# When researchers pay to publish: Results from a survey on APCs in four countries[1]

Osvaldo Gallardo,[1] Matías Milia,[2] André Luiz Appel,[3] Grip-APC Team[4] and François van Schalkwyk[5]

1 Consejo Nacional de Investigaciones Científicas y Técnicas (CONICET); Facultad de Ciencias Políticas y Sociales, Universidad Nacional de Cuyo (UNCuyo); Argentina. https://orcid.org/0000-0003-0662-2196
2 Department of Anthropology, University Notre Dame, United States. https://orcid.org/0000-0001-8474-5373
3 Instituto Brasileiro de Informação em Ciência e Tecnologia, Brazil. https://orcid.org/0000-0002-9608-803X
4 Université Paris Cité, France. https://orcid.org/0009-0000-5418-9935
5 Centre for Research on Evaluation, Science and Technology (CREST), Stellenbosch University, Stellenbosch, South Africa. https://orcid.org/0000-0002-1048-0429

**Abstract**
This paper provides an empirical overview of the impact and practices of paying Article Processing Charges (APCs) by four nationally categorized groups of researchers in Argentina, Brazil, Mexico, and South Africa. The data was collected from 13,577 researchers through an online questionnaire. The analysis compares the practice of publishing in journals that charge APCs across different dimensions, including country, discipline, gender, and age of the researchers. The paper also focuses on the maximum amount APC paid and the methods and strategies researchers use to cover APC payments, such as waivers, research project funds, payment by coauthors, and the option to publish in closed access, where possible. Different tendencies were identified among the different disciplines and the national systems examined. Findings show that Argentine researchers apply for waivers most frequently and often use personal funds or international coauthors for APCs, with younger researchers less involved in APC payments. In contrast, Brazil, South Africa, and Mexico have more older researchers, yet younger researchers still publish more in APC journals. South African researchers lead in APC publications, likely due to better funding access and "read and publish" agreements. This study lays the groundwork for further analysis of gender asymmetries, funding access, and views on the commercial Open Access model of scientific dissemination.

---

[1] This research had the financial support of the Global Research Institute of Paris, Idex Université Paris Cité, France, ANR-18-IDEX-0001.

**Resumen**

Este artículo presenta un análisis empírico del impacto y de las prácticas en torno al pago de *Article Processing Charges* (cargos por procesamiento de artículos o APC) en cuatro grupos de investigadores/as categorizados/as en Argentina, Brasil, México y Sudáfrica. Los datos fueron obtenidos a través de un cuestionario online que cosechó 13.577 respuestas. Se analiza comparativamente el impacto de la publicación en revistas científicas con APC a través de dimensiones como el país, la disciplina, el género y la edad de los/as investigadores/as. El trabajo también observa los montos máximos reconocidos en términos de APC y los distintos métodos para hacerles frente, como el pedido de descuentos o exenciones (*waivers*), fondos de proyectos de investigación, el pago a través de coautores/as y la opción por no publicar en Acceso Abierto, donde esta estuviera disponible. Distintas tendencias fueron identificadas a través de las disciplinas y los países. En Argentina. El análisis revela que los/as investigadores/as de Argentina solicitan *waivers* y pagan a través de colegas de otros países más frecuentemente que en los otros casos, además de que el grupo más joven declara publicar menos frecuentemente en revistas con APC. En contraste, en Brasil, Sudáfrica y México son los/as investigadores/as más jóvenes quienes publican más en ese tipo de revistas. La proporción de autores/as publicando en revistas con APC es máxima en el grupo sudafricano, lo que probablemente esté vinculado con mejor acceso a financiamiento y a la existencia de acuerdos *read and plublish*. El estudio provee un panorama general a partir del cual se conducirán análisis más específicos sobre asimetrías de género, acceso a financiamiento y puntos de vista sobre el modelo de comunicación científica basada en el Acceso Abierto comercial.

## Introduction

When the Budapest Open Access Initiative (BOAI) was launched in 2002, journal funding was one of the issues that came up for discussion, as the declaration promoted immediate and free access to scholarly publications. It was argued then that digital publishing was significantly cheaper than print. However, the BOAI did not advocate any particular publishing model to ensure the sustainability of open accessible scholarly journals (Budapest Open Access Initiative, 2002).

Two decades later, the picture is certainly very different. In the 2022 Declaration, two of the four sets of recommendations deal with article processing charges (APCs), one directly and one in the form of read and publish agreements (Budapest Open Access Initiative, 2022). Both statements emphasize that not all solutions are likely to work equally well in all contexts, thus highlighting the asymmetries between scholars and science systems across the world.

The 21st century has seen important changes in the political economy of scientific knowledge. With respect to journals, open access (OA) publishing models have enhanced internet-based accessibility to research results, rather than being restricted by paywalls typical of traditional subscription-based print publishing. This allows for broader and more inclusive dissemination and visibility of scientific knowledge (Piwowar et al., 2018; Tennant et al., 2016). However, the OA era coexists with the commoditization of scientific knowledge in the form of publications (Guédon & Loute, 2017). The scientific communication "game" now involves not only scientific institutions but also, and perhaps above all, oligopolistic for-profit firms (Larivière et al., 2015).

For its part, Latin America is a pioneering region in non-commercial OA, with journals supported by university and public scientific institutions, the vast majority of which do not charge APCs (Beigel, Sánchez Pereyra, et al., 2023; Beigel et al., 2024). In Africa, arguments have been made that the gold OA model, which relies on APCs, creates an economic barrier that limits the circulation of African-published research,while more equitable green OA models are not yet able to attract African researchers (Van Schalkwyk, 2024).

To complicate things further, in the Global South, authors struggle to find financial support to pay for APCs in their host institutions, getting financial aid from research funders and even from funding sources specially created to support OA publishing (Cantrell & Swanson, 2020; De-Castro & Franck, 2019). Full or partial waivers offered by journal editors shine, then, as an attractive yet uncertain alternative when there are no other ways of covering those costs.

Based on bibliometric data, some studies have estimated the possible maximum cost of APC payments. For Colombia, a total cost of USD 10,900,808 was estimated for the period 2009-2019. If only articles with corresponding authors from Colombian institutions are considered, the amount is USD 2,986,510 (Vélez Cuartas et al., 2020). For Chile, a maximum expenditure of USD 9,129,939 has been estimated for 2019 alone (Krauskopf, 2021). For authors in Uruguay, a quadrupling of APC expenditures was estimated between 2016 and 2019 (Tosar, 2022). For Argentina, a total of USD 31,059,865 was estimated for the period 2013-2020 (USD 11,634,112 for articles with Argentine researchers and corresponding authors) (Vélez Cuartas et al., 2022).

In this context, this article presents the results of a study on the impact of APCs in four peripheral countries – Argentina, Brazil, Mexico and South Africa – hosted by the Global Research Institute of Paris (GRIP), Université Paris Cité. The study involved the design and implementation of a survey that was conducted in the four countries in the second half of 2023. The target populations in these two continents share similar structural constraints, although they have some differences in internationalization and the weight of the evaluation systems in publishing practices. The survey allows for the analysis of the APC model and the problems that this model presents for researchers from non-hegemonic countries.

For some of the four countries, there are a limited number of related studies. In the case of Brazil, the journals published in the country that charge APCs are few (only 6% of those indexed in DOAJ in 2018) and the average APC values were lower than in the core countries. However, the values were higher for agricultural sciences and medical sciences journals as these fields have moved towards the APC OA model (Appel & Albagli, 2019). A bibliometric study projected that the average APC for the period 2012-2016 was close to USD 1,000 (Pavan & Barbosa, 2018).

In Argentina, a study of researchers at CONICET, the country's main scientific institution, observed in 2020 that the two most important criteria for choosing a journal were impact and no fees or the absence of APCs. The latter was even more important for researchers in agricultural, engineering, biological, and medical sciences (Zukerfeld et al., 2023). Another study conducted by Argentina's main research funding agency (the *Agencia I+d+i*) found that those who had used funds from their projects to pay APCs were overwhelmingly from the biological and medical sciences (Beigel & Gallardo, 2022). Finally, a bibliometric study showed that the projected APC expenditure for articles with CONICET authors increased by 45% in the period 2013-2020 (Vélez Cuartas et al., 2022).

In South Africa, Fullard's (2013) survey of 163 biomedicine researchers found that "the notion of 'paying to publish' was not seen as an insurmountable problem". More recently, Raju et al. (2017) report from one of the leading research universities in the country that the university has increased its APC budget, despite a general climate of fiscal austerity.

Recent studies show that authors from the Global South are underrepresented in the publication of OA articles (Borrego, 2023; Frank et al., 2023; Smith et al., 2022), suggesting that APCs remain a significant obstacle for these researchers. This stresses the importance of perspectives that address this question, considering regional and scientific periphery implications of the ongoing corporate capture of the OA through APCs.

This article focuses on researchers' practices in the context of the growing number of journals adopting the commercial OA model with APCs. The first section presents the target populations, including a descriptive statistical analysis of the samples obtained. After the research design, we present the results in three sections. The first section demonstrates the impact of publications in APC journals over the last five years, with analysis segmented by scientific discipline, gender, and age. Subsequently, we present the means used to pay APCs and the maximum amounts paid. Finally, survey results show the relationship between APC payments and international collaboration.

## Methodology

The study was designed as a national comparative study. The research team collaborated to design a common survey instrument, with minor variations to accommodate how the science system works in each national context and language issues. The questionnaire was finalized in English before being translated into Spanish and Portuguese.

Each national team was asked to write up a detailed description of the proposed target population consisting of scientific researchers in each country. This was done to ensure, to the extent possible, comparability across the four samples. The information regarding the national samples is provided below.

### The target populations

The historical background and contemporary contexts of the countries included in this study are diverse. Nevertheless, the research process attempted to ensure the selection of specific comparable populations of academics within each of the national scientific research environments.

In the case of Brazil, the institution selected to represent the population of researchers to be surveyed was the *Conselho Nacional de Desenvolvimento Científico e Tecnológico* (National Council for Scientific and Technological Development), commonly known as CNPq. The CNPq was created in 1951 and is one of the main Brazilian research funding bodies, with public funds provided by the Ministry of Science, Technology, and Innovation. Among its mandates are promoting scientific, technological and innovation research, and promoting the formation of qualified human resources for research in all areas of knowledge.

To that end, CNPq provides research grants to students in training (scientific initiation grants, master's, doctorate, postdoctoral, in different modalities) and to researchers affiliated with Brazilian teaching and research institutions (called “research productivity grants” or “*Bolsa Produtividade em Pesquisa*” – PQ). These, in turn, are awarded to prominent researchers in their areas of expertise, and who are evaluated by Advisory Committees that consider, among other criteria, the number of articles published in journals indexed in bibliographic databases, the Impact Factor of these publications and the performance of the applicants related to the training of human resources for research and innovation.

The PQ system is organized into eight major scientific areas (see Table 1) and tiers (in ascending order: 2, 1D, 1C, 1B, 1A, with the last four tiers collectively referred to as “tier 1”). Each level provides scholarships of different amounts. In addition to the scholarship, there are some advantages in the case of tier 1, such as exclusive funding calls for Tier 1 researchers and an additional monthly payment for research-related costs. In addition, only Tier 1 researchers can be members of the evaluating committees.

*Table 1. Brazil's target population (researchers holding CNPq research productivity grants) by major scientific field, 2024. N=15,084*

| Major scientific area | n | % |
|---|---|---|
| Agricultural sciences | 1,952 | 13% |
| Biology sciences | 2,452 | 16% |
| Health sciences | 1,652 | 11% |
| Science and earth sciences | 3,218 | 21% |
| Humanities | 1,984 | 13% |
| Applied social sciences | 1,170 | 8% |
| Engineering | 2,034 | 13% |
| Language, linguistics, and arts | 622 | 4% |
| **Total** | **15,084** | **100%** |

Source: Authors' elaboration based on data from http://memoria2.cnpq.br/web/guest/bolsistas-vigentes

In Mexico, the institution selected was the *Sistema Nacional de Investigadores e Investigadoras* (SNII, National System of Researchers), part of the *Consejo Nacional de Humanidades, Ciencia y Tecnología* (CONAHCYT). It was established in 1984 to promote research and in response to "brain drain", that is, talent leaving the country for better economic opportunities elsewhere in the world during Mexico's foreign debt crisis (Galaz-Fontes & Gil-Antón, 2013; Gil-Antón, 1996).

Membership at the SNII is prestigious, especially at the highest levels, which requires demonstrating significant contributions in research as well as teaching and the promotion of research in Mexico. Evaluation for renewal and admission of new members relies mostly on their academic credentials and production, but also includes other factors such as the creation of programs and projects. Research assessment is based on peer-review and reviewers are partially chosen by members of the community and appointed by the SNII.

To become and remain a member of the SNII, researchers must show a history of systematic research in their fields. According to the last calls for admission and promotion, applicants do not need to have full-time employment with a university or research institution, although this is still the case for most applicants. Researchers are evaluated through the publication of articles, books, research results, and book chapters as well as patents, technological innovations and technology transfer, if applicable. Other areas for evaluation include the supervision of doctoral candidates, the teaching of undergraduate and graduate courses, and the development of new researchers and the establishment of research groups.

There are five tiers in SNII. For those who are starting their research careers, SNII grants an introductory level of membership, that is, "Candidate". For more established researchers, the levels are Level I, Level II, Level III, and National Researcher Emeritus (*Investigador Nacional Emérito*). To be admitted as members, applicants must hold a doctoral degree, demonstrate the ability to perform original research in their fields, and must have finished their bachelor's degree not more than 15 years prior to the application date (although exceptions are made in the case of this requirement). For Level I, the researcher must also show work in the supervision of theses, and the teaching of undergraduate and graduate courses, as well as evidence of the creation of knowledge in their fields. For Level II, in addition to the previous requirements, a Level II member's research outputs must be recognized as having made a leading contribution in the field with its quality being assessed by senior members of the SSII. Level III is reserved for those researchers who have made significant contributions to their fields in Mexico and who have obtained international recognition. All levels are subject to periodic reevaluation, except for National Researcher Emeritus, which is for life.

Based on their SNII level, researchers receive a cash bonus that tends to complement their salary. For some, this bonus is between one-third to half of their monthly income, and the SSII bonus is therefore an essential component of many researchers' incomes. According to CONAHCYT, in the 1st quarter of 2023, the SNII counted 41,367 active members.

*Table 2. Mexico's target population (researchers categorised by SNII), by major scientific field, 2024. N=41,367*

| **Major scientific area** | **N** | **%** |
|---|---|---|
| Physics, mathematics and earth sciences | 5,373 | 13% |
| Biology and chemistry | 6,678 | 16.1% |
| Medicine and health sciences | 4,583 | 11.1% |
| Behavioral and educational sciences | 815 | 2% |
| Humanities | 5,565 | 13.5% |
| Social sciences | 7,512 | 18.2% |
| Agriculture, livestock, forestry, and ecosystem sciences | 4,196 | 10.1% |
| Engineering and technological development | 6,234 | 15.1% |
| Interdisciplinary | 411 | 1% |
| **Total** | **41,367** | **100%** |

Source: Authors' elaboration based on data from Archivo Histórico del SNII (see https://conahcyt.mx/sistema-nacional-de-investigadores/archivo-historico/).

For Argentina, the institution selected was the *Consejo Nacional de Investigaciones Científicas y Técnicas* (CONICET, National Scientific and Technical Research Council) created in 1958. At first, it was structured around the natural and biological sciences by Bernardo Houssay (the first Argentinian Nobel Prize winner in medicine) its first president. CONICET was created following the model of the French CNRS. Currently, it includes researchers, doctoral and postdoctoral fellows, and a national network of research institutes from all scientific areas. CONICET is a research agency employing full-time researchers; it is not a higher education institution as it does not include teaching in undergraduate or postgraduate programmes. CONICET is the main funding source for doctoral and postdoctoral scholarships awarded to fulfill a PhD in Argentina.

CONICET research institutes can be exclusively dependent on the national agency or have an institutional partner – frequently a national university – which shares its management and funding. CONICET offers tenured positions for researchers which are completely independent from a position at a university. Teaching positions are acquired through university positions and based on university regulations. It is frequent that CONICET researchers hold part-time teaching positions.

Conditions for entrance to CONICET are similar to those for a tenure position at a university including academic background, publication record and research projects. Once a year, there is a public call for admission and candidates can present themselves to one of the five categories available: Assistant, Associate, Independent, Principal, and Superior (Senior). Normally, young scholars aspire to be in the first category (Assistant). The candidates are evaluated by disciplinary committees composed of peers who also ask for expert advice to assess applications.

The progress across the CONICET career categories is mainly linked to publishing productivity and the Impact Factor of the journals. Other aspects are also considered, depending on the category (internationalization, leading role in publications, teaching positions, supervision of PhD candidates, participation in and/or leading research projects, technology transfer, etc.). The natural and exact sciences have developed an extensive tradition of internationalization and mainstream publishing, using bibliometric indicators and author position as the most relevant criteria for career-building. Differently, the social sciences and humanities have specific regulations at CONICET for classifying their journals: bibliometric indicators are scarcely used, and journals indexed in Latin American services are highly valued.

CONICET admission vacancies vary between 600 and 900 per annum and are equally distributed in the traditional four major scientific areas. As a result of this policy, the four major areas currently have a similar number of researchers, while he recently created fifth major area, "technology" is still very small.

*Table 3. Argentina's target population (full-time researchers of CONICET), by major scientific field, 2024. N=12,176*

| Major scientific area | n | % |
|---|---|---|
| Engineering and agricultural sciences | 3,313 | 24.7% |
| Biological and health sciences | 3,287 | 27% |
| Natural and exact sciences | 2,580 | 21.2% |
| Social sciences and humanities | 2,977 | 24.4% |
| Technology | 327 | 2.6% |
| **Total** | **12,176** | **100%** |

Source: Authors' elaboration based on data from https://cifras.conicet.gov.ar/publica/

South Africa's National Research Foundation (NRF) is the funding agency of the Department of Science and Innovation of the government of South Africa. The NRF supports research and innovation; encourages an interest in science and technology; and facilitates high-end research. It provides funding to the 26 country's public universities, as well as to science centers, research facilities and other science organizations. The NRF was established on 1 April 1999 as an autonomous statutory body in accordance with the National Research Foundation Act.

Researchers in South Africa voluntarily elect to be rated by the NRF. They do so for the purposes of peer recognition as well as to improve their chances of promotion, and of being funded by the NRF.

The categorization system used by the NRF to rate researchers in South Africa allows the population to be disaggregated by level of international collaboration and reputation, by career progression (emerging or established researcher), and by institutional affiliation. The ratings consist of the following five categories:

1. A (with sub-designations A1 and A2) for leading international researchers
2. B (with sub-designations B1, B2 and B3) for internationally recognized researchers
3. C (with sub-designations C1, C2 and C3) for established researchers
4. P (prestigious award) for emerging scholars under 35 years of age
5. Y (with sub-designations Y1 and Y2) for promising young researchers under 40 years of age.

The first three (A to C) apply to established researchers and the last two categories (P and Y) apply only to young and emerging researchers. Academics over the age of 40 do not qualify for the Y category and those older than 35 years do not qualify for the P category. Additionally, even if candidates are younger than the age categories for these ratings but graduated with a PhD more than five years previously, they also do not qualify for the emerging researcher category. Those without a PhD degree do not qualify for the emerging researcher categories.

Those seeking an A, B or C rating will have their applications considered based on research output. This is not only based on the number of publications or research outputs produced, but on the answers to questions in the evaluation forms completed by peers who are asked to review applications, including what their opinions are of the applicant, the applicant's research, and the impact of the applicant's research.

The NRF captures data on each rated researcher's primary and secondary field of study. In the case of their primary field, researchers must select from the NRF's own field categories and are permitted to select more than one primary field. Table 4 shows NRF-rated researchers by major field of study using only the first field chosen in cases where more than one field was selected.

*Table 4. South Africa's target population (researchers categorized by NRF), by major scientific field, 2024. N=4,391*

| Major scientific area | n | % |
|---|---|---|
| Biological sciences | 570 | 13.0% |
| Social sciences | 564 | 12.8% |
| Humanities | 512 | 11.7% |
| Health sciences | 379 | 8.6% |
| Engineering sciences | 294 | 6.7% |
| Physical sciences | 285 | 6.5% |
| Agricultural sciences | 207 | 4.7% |
| Mathematical sciences | 206 | 4.7% |
| Chemical sciences | 200 | 4.6% |
| Earth and marine sciences | 192 | 4.4% |
| Economic sciences | 192 | 4.4% |
| Medical sciences: Basic | 152 | 3.5% |
| Technologies and applied sciences | 150 | 3.4% |
| Information and computer science | 144 | 3.3% |
| Law | 132 | 3.0% |
| Medical sciences: Clinical | 113 | 2.6% |
| Arts | 99 | 2.3% |
| **Total** | **4,391** | **100%** |

Source: Authors' elaboration based on data from the NRF's website.

In summary, there are several common characteristics among the evaluation cultures of the four cases under study. They consist of nationally funded systems for non-binding research categorization. Bibliometric indicators play a significant role in entry and career advancement, with the highest tiers in journal rankings being associated with substantial international impact in terms of citations. In the four national cases local impact is not highly prioritized, and research classification and tenure in teaching are not necessarily interconnected.

In terms of distinctions, Mexico emphasizes the significance of salary bonuses and holds the largest number of researchers within its system. Meanwhile, in Argentina, the target population holds full-time research positions in CONICET, a feature not present in the three other countries. Within CONICET, salary discrepancies between different categories are minimal and not attached to publishing performance. Interestingly, social sciences and humanities researchers make up a third of the total in South Africa and Mexico, and a quarter in Brazil and Argentina.

### Data collection instrument

An online questionnaire (see Annex) was designed to cover various research dimensions, including the following in five sections: (1) general information about position, age, gender, disciplines, and research activities; (2) publication of results and evaluation of journals; (3) views and practices on open access scholarly publishing; (4) views and practices on paying APCs and on journal publication; and

(5) challenges regarding APCs and open questions. This article focuses only on sections 1 and 4 of the questionnaire.

The questionnaire was translated from English into Spanish and Portuguese. All four questionnaires were identical, except for the questions in Section 1 which related to the researchers' rating or membership level at the scientific organization in their respective country. The survey instrument was an online questionnaire set up using LimeSurvey software. Following a piloting process in which 10 colleagues in each country were asked to complete the online survey, and to provide written feedback on any issues encountered. The questionnaires were revised based on the feedback received, and the revised and final questionnaires were sent out at various times between September and November 2023. The last closing date for submitting questionnaires was on 15 February 2024 (South Africa).

The research project, its survey design and instruments were approved by the Research Ethics Committee: Social, Behavioural and Education Research (SBER) at Stellenbosch University (Project number: CREST-2023-29324).

***Argentina***: A link to the online questionnaire was sent formally by CONICET via email to the 12,176 CONICET full-time researchers included in the institutional database, on 27 September 2023. In total, 3,313 complete responses were received. Incomplete questionnaires or respondents who stated that they were not CONICET researchers were discarded. Two email reminders were sent, and the last response was received on 24 November 2023.

***Brazil***: The online questionnaire was sent to 15,426 researchers categorized by the CNPq (*bolsistas de produtividade*) on 30 August 2023. The survey was closed on 30 September. A total of 6,288 full responses were received. The invitation email to participate in the survey was sent to the researchers by the official institution selected, CNPq.

***South Africa***: The link to the online questionnaire was sent via email by the project team to 3,808 NRF-rated researchers for whom email addresses were available on 27 November 2023. Email addresses were obtained by cross-referencing the publicly available list of NRF-rated researchers with the SA Knowledge base of the Centre for Research on Evaluation, Science and Technology (CREST) at Stellenbosch University. SA Knowledge base is a database of all publications in South Africa that qualified for financially subsidy as part of the South African government's program for rewarding public universities for publications. Specifically, name and affiliations were compared to find matches with corresponding authors in the SA Knowledge base. A total of 547 completed surveys were received; 535 complete responses were retained after 12 questionnaires were discarded because the respondents' NRF ratings had lapsed. Three reminders were sent, and the survey was closed on 15 February 2024.

***Mexico:*** The survey was sent to 23,078 SNII researchers, which was the number of email addresses obtained from bibliometric sources (based on a search conducted in Scopus). The online survey was available from 11 October to 28 December 2023, and a reminder was sent a month after the original survey. A total of 3,441 complete responses were retained for the statistical analysis.

Table 5 shows the responses obtained from the four countries. In the national cases where the survey was sent by the official institutions involved in the study (Brazil and Argentina), the response rates significantly were higher. In contrast, the rates were lower for the two countries (Mexico and South Africa) for which the invitation emails were sent out by the research team. In addition, it is important to note that in both South Africa and Mexico surveys, the population that received the invitation was lower than the original target population given the limitations of the mailing database (see Tables 2 and 4). The differences are due to the strategies of email address collection (also performed by the authors)[2].

*Table 5. Target population and valid responses obtained, by country*

| Country | Rating agency | Target population | Valid responses obtained | Sample size |
|---|---|---|---|---|
| Argentina | CONICET | 12,176 | 3,313 | 27.2% |
| Brazil | CNPq | 15,426 | 6,288 | 40.8% |
| Mexico | SNII | 41,367 | 3,441 | 8.3% |
| South Africa | NRF | 4,391 | 535 | 12.2% |

Source: Authors' elaboration based on data from APC surveys 2023 (GRIP project).

## Findings and discussion

### Descriptive statistics of the national samples

The distribution of researchers across scientific fields is quite similar in all four countries (refer to Table 6). When compared to the distributions presented in Tables 1-4, the scientific field distributions of the samples closely resemble those of the target population. The samples were not built through probabilistic procedures and, therefore, statistically valid inferences cannot be made; nevertheless, they are still high-quality samples, to the extent that they represent relevant proportions of carefully selected populations of researchers in each of the four countries.

Comparing the morphology of each national sample we see in Table 6 that the share of researchers in Biological Sciences is higher in Argentina compared to Mexico, Brazil and South Africa, while the opposite is true for Medical and Health Sciences. South Africa displays one unique characteristic with Social Sciences and Humanities scholars being the most highly represented group compared with the other three countries. In addition, the proportions of South African researchers who selected "Two different scientific areas", along with those who selected Agricultural Sciences, and Engineering and Technology, are lower compared with the other countries.

[2] In Brazil and Argentina, the team received valuable assistance from CNPq and CONICET, respectively, in reaching out to the target population for the survey. In South Africa, email addresses were obtained from the SA Knowledge database and helped the research team to reach a reasonable proportion of the target population. However, in the case of Mexico, neither of these methods could be used. Departing from the 41,367 researchers included in the public census published by the SNII (see https://conahcyt.mx/sistema-nacional-de-investigadores/archivo-historico/), emails were collected using a bibliometric approach based on publications by Mexican-affiliated authors indexed in Scopus. Only 23,078 email addresses were successfully collected.

*Table 6. Researchers by major scientific field and country. N=13,577.*

| Major field | Argentina | Mexico | Brazil | South Africa |
|---|---|---|---|---|
| Agricultural Sciences | 6,0% | 8,5% | 8,5% | 3,7% |
| Basic and Natural Sciences | 18,4% | 13,0% | 16,7% | 15,1% |
| Biological Sciences | 17,2% | 13,6% | 10,2% | 12,5% |
| Engineering and Technology | 11,1% | 11,5% | 9,9% | 7,1% |
| Medical and Health Sciences | 6,5% | 11,9% | 9,8% | 12,1% |
| Social Sciences and Humanities | 21,8% | 23,2% | 24,0% | 36,8% |
| Other* | - | 2,2% | - | - |
| Two major scientific areas | 19,1% | 16,1% | 20,9% | 12,5% |
| **Total** | **100%** | **100%** | **100%** | **100%** |

Source: Authors' elaboration based on data from APC surveys 2023 (GRIP project).
Note: only the SNII-Mexico survey included the option "Other".

The survey indicates strong representation in both the Biological and Medical and Health Sciences; these are scientific fields whose researchers usually pay APCs more frequently (Nwagwu, 2023; Pavan & Barbosa, 2018; Solomon & Björk, 2012). As mentioned, South Africa stands out because most respondents were from the Social Sciences and Humanities. This is worth noting because it may reflect a growing concern among researchers in this field as they are more frequently confronted with having to pay APCs or finding financial support for paying APCs in a research environment where such support is scarce. On the other hand, South Africa is the country in the sample with a relatively large number of read and publishing deals which cover more than 9,000 journals from various publishers.[3]

In terms of gender, three of the national samples have a predominance of male researchers (see Table 7). In Argentina, respondents were mostly female (56,2%) in a similar proportion with the gender composition of the universe at CONICET; however male researchers are more highly represented in the more prestigious categories such as Principal and Superior (Beigel, Almeida, et al., 2023; Gallardo, 2022), similar to the situation in other countries.

[3] See https://sanlic.ac.za/read-and-publish-agreements/#rp_agreements

*Table 7. Researchers by gender and country. N=13,577*

| Gender | Argentina | Mexico | Brazil | South Africa |
|---|---|---|---|---|
| Male | 42,4% | 60,1% | 63,7% | 57,4% |
| Female | 56,2% | 38,8% | 35,5% | 41,5% |
| Non-binary | 0,3% | 0,1% | 0,1% | 0,0% |
| Other | 0,0% | 0,0% | 0,1% | 0,0% |
| Prefer not to say | 1,1% | 0,9% | 0,6% | 1,1% |
| **Total** | **100%** | **100%** | **100%** | **100%** |

Source: Authors' elaboration based on data from APC surveys 2023 (GRIP project).

The gender balance varies more by major scientific area (see Table 8). In all four countries, the proportion of women researchers is highest in Medical and Health and in Social Sciences and Humanities. In Argentina, the proportion of women is also higher in Agricultural Sciences. Engineering and Technology, and Basic and Natural Sciences are the areas with the lowest proportion of female researchers.

*Table 8. Percentage of female researchers in each major scientific area, by country. N=13,577*

| Major area | Argentina | Mexico | Brazil | South Africa |
|---|---|---|---|---|
| Agricultural Sciences | 63,5% | 28,8% | 26,5% | 30,0% |
| Basic and Natural Sciences | 43,9% | 26,1% | 18,4% | 27,2% |
| Biological Sciences | 56,1% | 39,7% | 39,6% | 35,8% |
| Engineering and Technology | 48,2% | 24,3% | 17,2% | 29,0% |
| Medical and Health Sciences | 67,4% | 48,3% | 49,4% | 66,2% |
| Social Sciences and Humanities | 60,2% | 49,3% | 49,4% | 43,7% |
| Other | - | 44,0% | - | - |
| Two major scientific areas | 62,3% | 41,2% | 37,2% | 44,8% |
| **Total** | **56,2%** | **38,8%** | **35,5%** | **41,5%** |

Source: Authors' elaboration based on data from APC surveys 2023 (GRIP project).

We calculated both the biological age of respondents as well as their academic aged based on number of years after they obtained their apex degree. For the survey sample, there was a high level of correspondence between the biological and academic age in each country. We therefore present here only the results on the academic age of respondents.

Argentina has a younger age structure compared with the academic ages of respondents from the other three countries (see Figure 1). Approximately two-thirds of CONICET researchers obtained their PhD degree in the last 15 years. Only Mexico's SNII shows a similar result (almost 50% obtained their apex degree in the past 15 years), while Brazil and South Africa have significant proportions of researchers whose highest degree was obtained more than 30 years ago.

*Figure 1. Researchers by academic age (in years) and country. N=13,577*

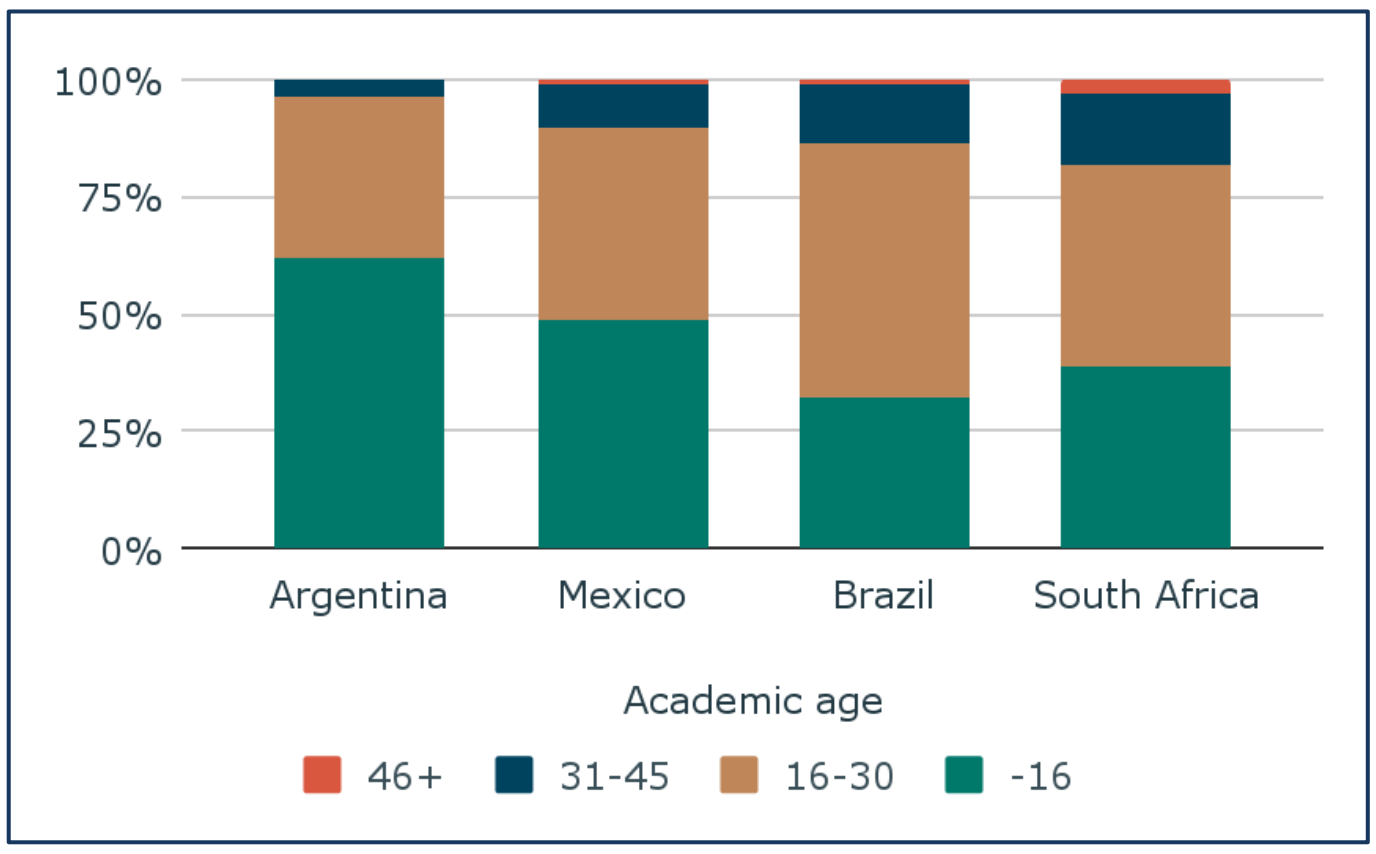

Source: Authors' elaboration based on data from APC surveys 2023 (GRIP project).
Note: Academic age bands have been calculated using 2023 as the base year and considering the year of the highest academic degree (in 99%, PhD).

The age profiles show similarity between the populations analyzed in Mexico, South Africa and Brazil, reflecting more senior researchers in the scientific systems of each country. In these countries, it is generally necessary to have an established academic track record to enter their respective scientific membership and rating systems. In CONICET, however, there exists a stepping-stone that makes possible the entry of younger researchers with a shorter academic career.

### Publishing in APC journals by discipline, age and gender

A large proportion of the researchers surveyed reported having published in journals that charge APCs at least once in the past five years. The share is highest among researchers from South Africa (77,4%) and lowest in Argentina (44,1%). The percentages for Mexico and Brazil are very similar, at around 66% (see Table 9). Publishing in a journal that charges APCs does not necessarily mean that the APC is actually paid by the responding researcher or, if it is paid, it can be a partial contribution. We will return to this topic below.

*Table 9. Percentage of researchers who have published in APC journals in each major scientific field, by country. N=13,577*

| Major field | Argentina | Mexico | Brazil | South Africa |
|---|---|---|---|---|
| Agricultural Sciences | 60,9% | 74,0% | 91,6% | 85,0% |
| Basic and Natural Sciences | 41,3% | 62,6% | 61,5% | 71,6% |
| Biological Sciences | 55,4% | 81,0% | 87,2% | 88,1% |
| Engineering and Technology | 38,8% | 73,2% | 70,4% | 84,2% |
| Medical and Health Sciences | 71,2% | 85,1% | 87,8% | 90,8% |
| Social Sciences and Humanities | 16,6% | 33,8% | 24,9% | 68,0% |
| Other | - | 73,3% | - | - |
| Two major scientific areas | 56,7% | 78,3% | 83,0% | 82,1% |
| **Total** | **44,1%** | **66,1%** | **65,9%** | **77,4%** |

Source: own elaboration based on data from APC surveys 2023 (GRIP project).

Table 9 shows the differences between the major scientific fields. Publishing in APC journals is more common for researchers in Agricultural, Biological, and Medical and Health Sciences, and for those who conduct their research in two major scientific fields. In all these fields, more than half of researchers have in the last five years published in a journal that charges APCs. Basic and Natural, and Engineering and Technology researchers do not report as a high a rate of publications which required the payment of APCs. Finally, the Social Sciences and Humanities researchers are those who typically have not paid APCs in the past five years.

It is important to note that the frequency of publications for which APCs have been paid is noticeably higher among South African researchers. In almost all cases, the corresponding values are higher than in the other countries, but the difference is much more pronounced comparing the Social Sciences and Humanities (almost three times as high as in Brazil and four times as high as in Argentina).

In terms of gender, 40% of female respondents report having published in APC journals in the last five years (compared with 49% of male respondents). Only in South Africa, is the proportion of women who have published in APC journals higher than that of men (see Table 10). In Mexico, Brazil and South Africa, the differences between men and women are noticeable in each major field but are generally small. In Argentina, however, the differences are greater. In the social sciences and humanities, for example, the percentage of men who have published in APC journals is almost double that of women. The higher incidence of APC journals among men than among women is found in virtually all segments defined by major disciplinary fields, career categories, and both biological and academic age.

*Table 10. Percentages of female and male researchers who have published in APC journals in each major scientific field, by country. N=13,577*

| Major field | Argentina | | Mexico | | Brazil | | South Africa | |
|---|---|---|---|---|---|---|---|---|
| | Male | Female | Male | Female | Male | Female | Male | Female |
| Agricultural Sciences | 68,6% | 56,8% | 76,6% | 69,1% | 92,8% | 88,7% | 85,7% | 83,3% |
| Basic and Natural Sciences | 43,2% | 39,3% | 62,3% | 64,1% | 60,1% | 67,9% | 72,9% | 68,2% |
| Biological Sciences | 64,5% | 48,9% | 85,3% | 74,7% | 88,3% | 85,4% | 85,4% | 91,7% |
| Engineering and Technology | 44,1% | 32,6% | 75,3% | 66,7% | 69,9% | 73,8% | 85,2% | 81,8% |
| Medical and Health Sciences | 75,0% | 69,7% | 87,0% | 83,3% | 86,8% | 88,9% | 86,4% | 93,0% |
| Social Sciences and Humanities | 21,2% | 13,8% | 35,1% | 32,3% | 25,3% | 24,4% | 67,3% | 67,4% |
| Other | - | - | 80,5% | 63,6% | - | - | - | - |
| Two major scientific areas | 66,4% | 51,3% | 78,1% | 79,4% | 83,8% | 81,6% | 75,7% | 90,0% |
| **Total** | **49,2%** | **40,4%** | **68,6%** | **62,2%** | **67,7%** | **62,9%** | **75,6%** | **79,3%** |

Source: Author's elaboration based on data from APC surveys 2023 (GRIP project).

In terms of academic age, early career researchers (i.e., those with less than 16 years since degree [-16] opposed to those between 16 and 30 years since degree) have a higher proportion of publications in APC journals (see Figure 2). The exception is Argentina, where the percentage is practically the same for each age group (there is only one respondent in the 46 years and older (46+) age band). The impact of the OA model with APCs therefore seems to be greater for the younger segments of the research community, particularly in Brazil and South Africa. In Mexico, this effect seems to be weaker. On the contrary, in Argentina, there seems to be no visible relationship between career stage and publication in APC journals.

*Figure 2. Researchers by country academic age band who have published in APC journals (% in years). N=13,577*

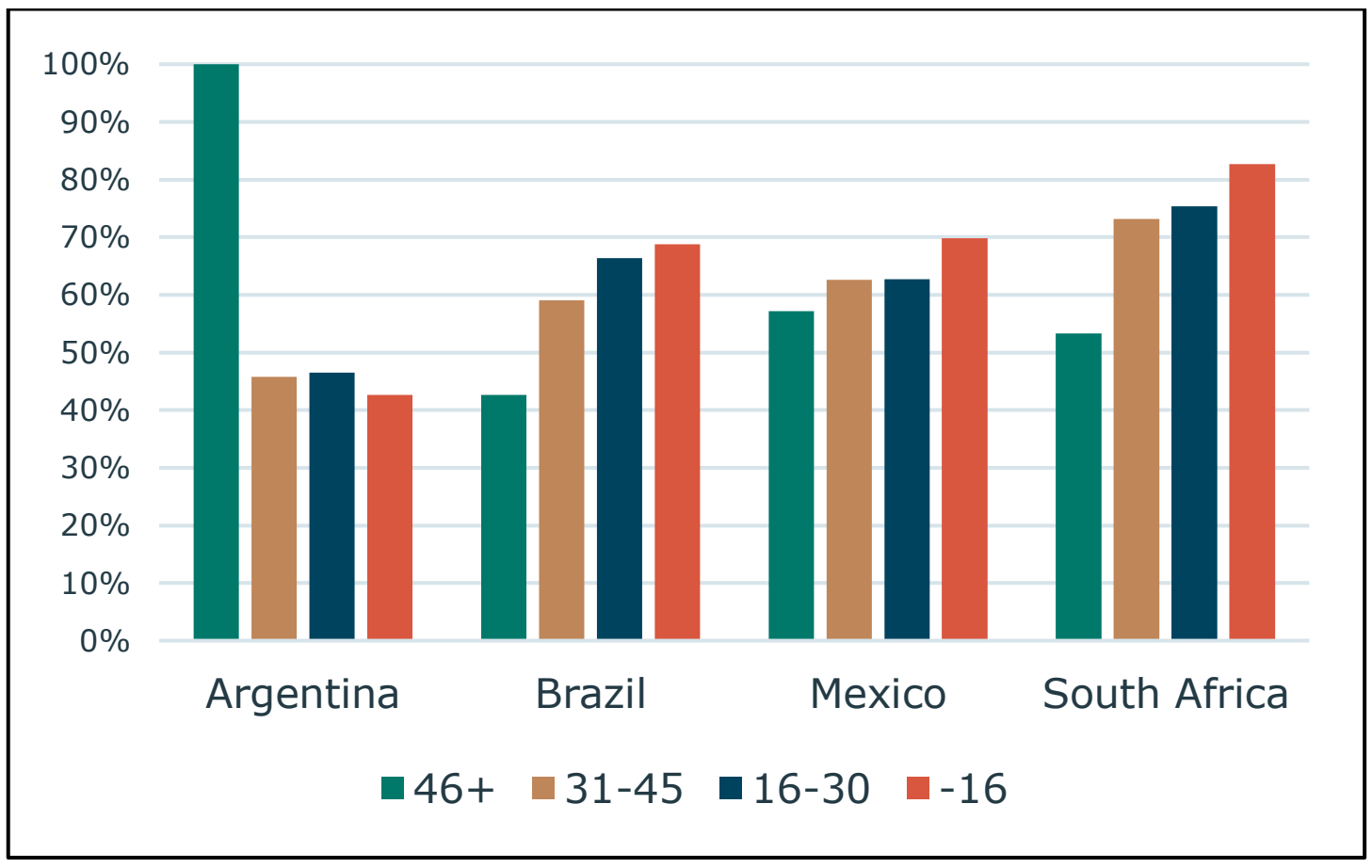

Source: Authors' elaboration based on data from APC surveys 2023 (GRIP project).
Note: Academic age bands have been calculated using 2023 as the base year and considering the year of the highest academic degree (in 99%, PhD).

## APC payment: methods and amounts

The modalities used to pay APCs vary from country to country (see Table 11). The data for Argentina reveals two notable characteristics. The first is that there appears to be little (2,1%) institutional-level support for researchers to pay APCs. By contrast, in South Africa, the second-most frequent modality is institutional funds.

The second notable characteristic for Argentina relates collaboration in the form of co-authorships. Of those who paid APCs, 17,4% of researchers in Argentina relied on international co-authors to pay APCs.

*Table 11. Most used modality for APC payment, by country. N=8,289*

| APC payment | Argentina | Mexico | Brazil | South Africa |
|---|---|---|---|---|
| Own research project funds | 24,7% | 23,5% | 29,4% | 48,1% |
| Personal funds | 5,5% | 15,8% | 34,5% | 1,9% |
| Partial or full waiver | 27,3% | 12,4% | 3,5% | 6,3% |
| No APC payment and closed access | 16,8% | 20,3% | 9,3% | 8,9% |
| Specific institutional APC funds | 2,1% | 11,8% | 13,1% | 24,6% |
| Payment by international co-authors | 17,4% | 7,7% | 3,2% | 5,6% |
| Payment by national co-authors | 3,7% | 8,5% | 6,0% | 2,9% |
| Other | 2,3% | 0% | 0,7% | 1,4% |
| Don't know | 0,3% | 0% | 0,3% | 0,2% |
| **Total** | **100%** | **100%** | **100%** | **100%** |

Source: Authors' elaboration based on data from APC surveys 2023 (GRIP project).

In Argentina, Mexico and Brazil "own research project funds" is one of the most frequent methods of paying for APCs, but only half as frequent as in the case of South Africa: in South Africa 48,1% of researchers relied on own research project funds to pay APCs, while 24,7% did so in Argentina, 23,5% in Mexico and 29,4% in Brazil. This is probably explained by South Africa's unique system of financial rewards paid to researchers which are to be used exclusively for research-related activities, including the payment of APCs. Mexico, comparatively speaking, is the country in which publishing articles without APCs is most common.

In terms of the maximum APCs paid, the range of USD 1,001 to 3,000 was the most frequently selected option in all four countries. South Africa stands out as the country in which researchers more frequently selected the category above USD 3,000 compared to the other three countries (see Figure 3-6). In South Africa, 18,8% of respondents paid APCs of more than USD 3,000. In the other three countries, the proportions of respondents who paid more than USD 3,000 in APCs were as follows: Brazil, 13,8%; Mexico, 12,8%, and Argentina, 8,0%.

In general, the more researchers in a country who pay APCs from institutional or own research funds, the higher the average APC paid in that country. Conversely, the more researchers from a country pay from own funds or rely on APC waivers, the lower the APCs paid, as is notable in the case of Argentina.

*Figure 3. Argentina: Researchers by major scientific field, maximum APC paid in USD. N=1,460*

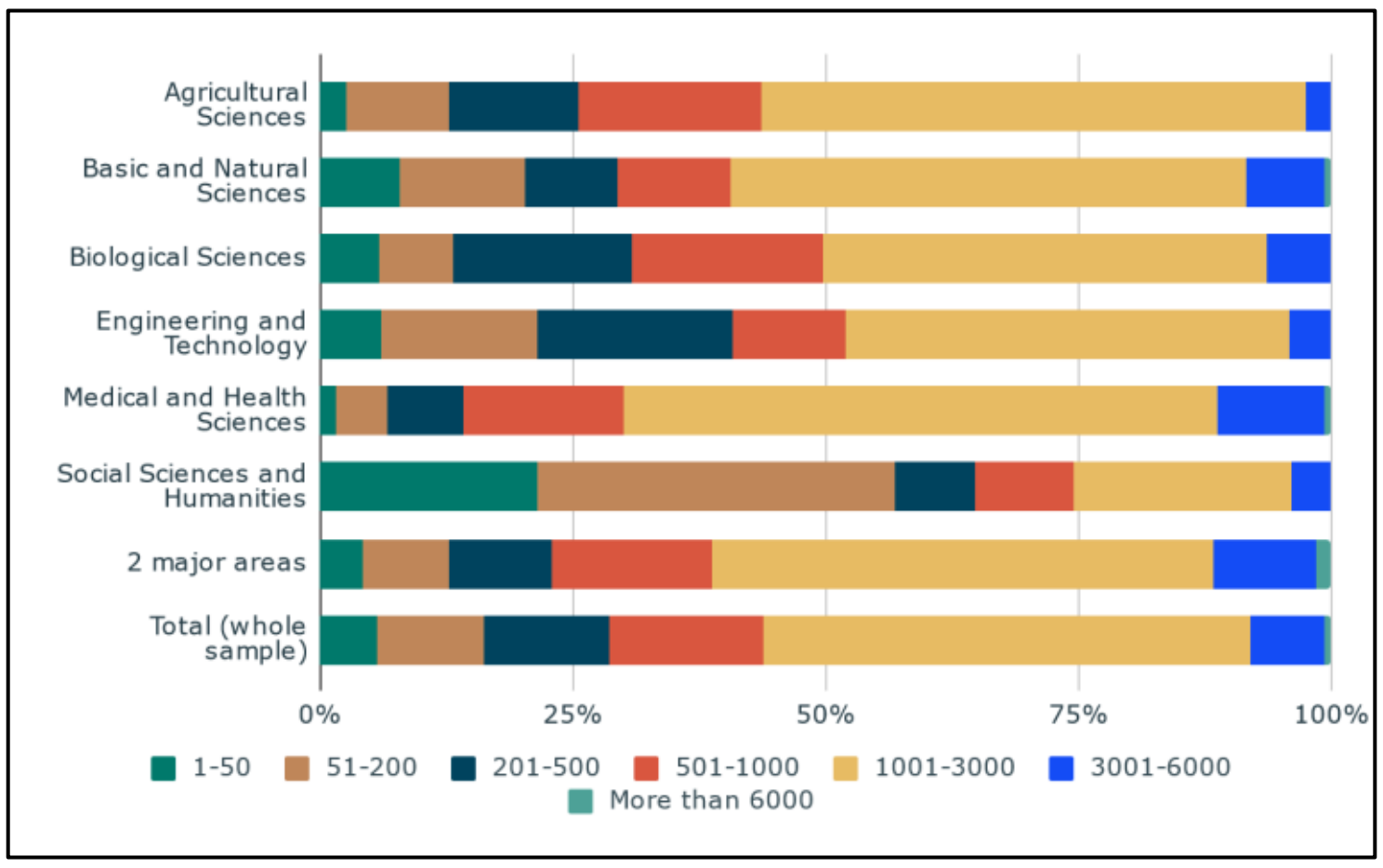

Source: Authors' elaboration based on data from APC surveys 2023 (GRIP project).

*Figure 4. Mexico: Researchers by major scientific field, maximum APC paid in USD. N=2,273*

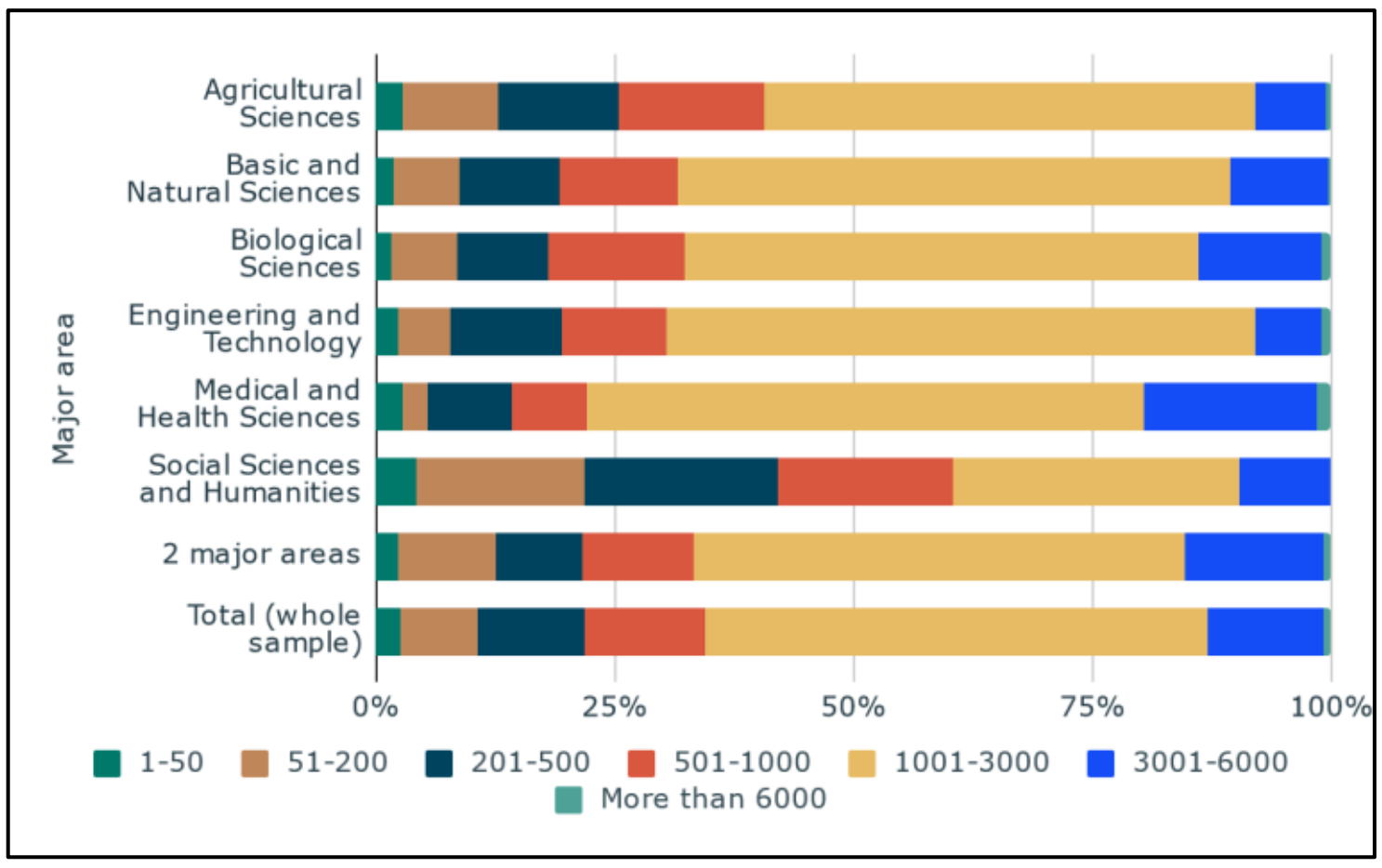

Source: Authors' elaboration based on data from APC surveys 2023 (GRIP project).

*Figure 5. Brazil: Researchers by major scientific field, maximum APC paid in USD. N=4,142*

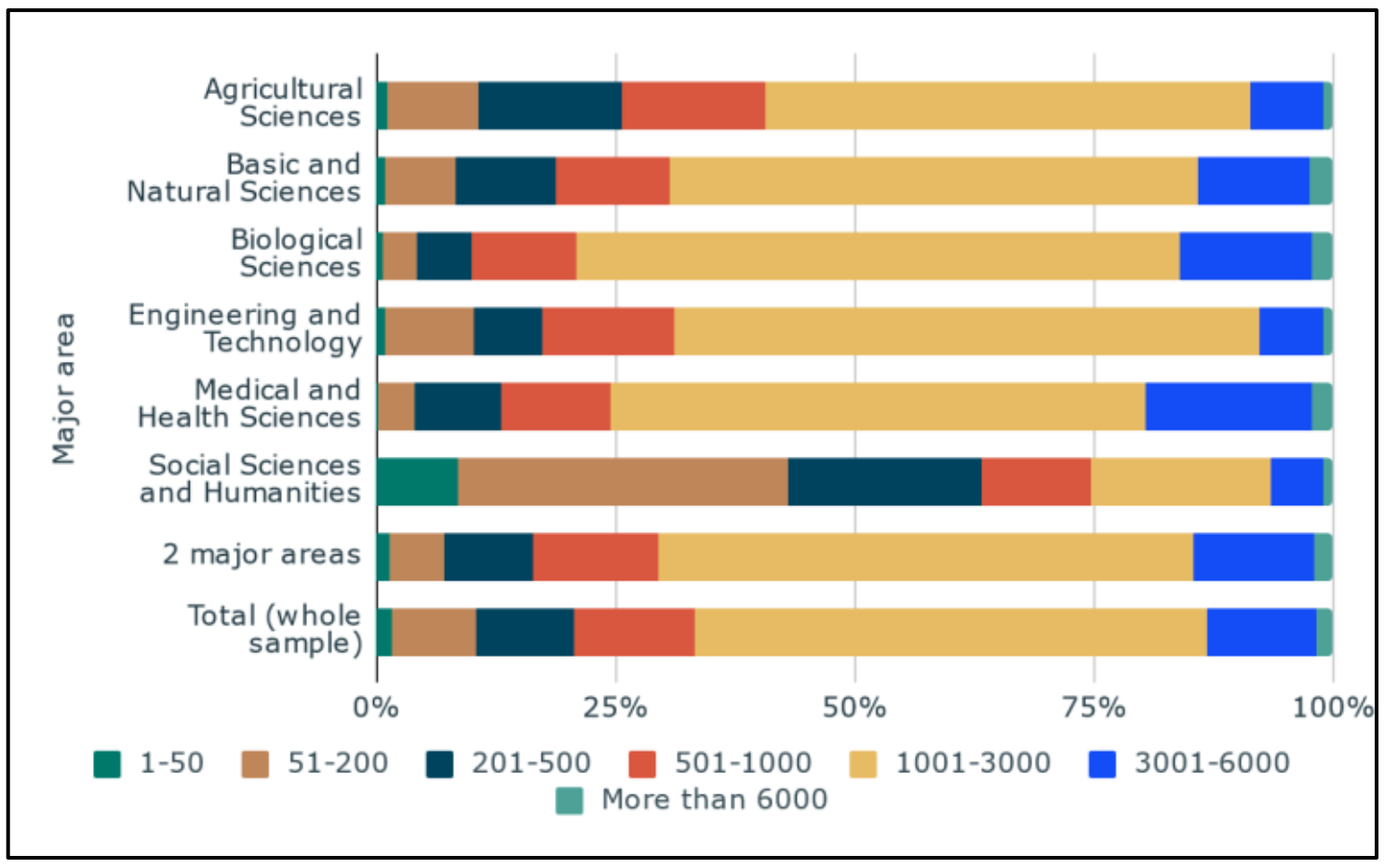

Source: Authors' elaboration based on data from APC surveys 2023 (GRIP project).

*Figure 6. South Africa: Researchers by major scientific field, maximum APC paid in USD. N=414*

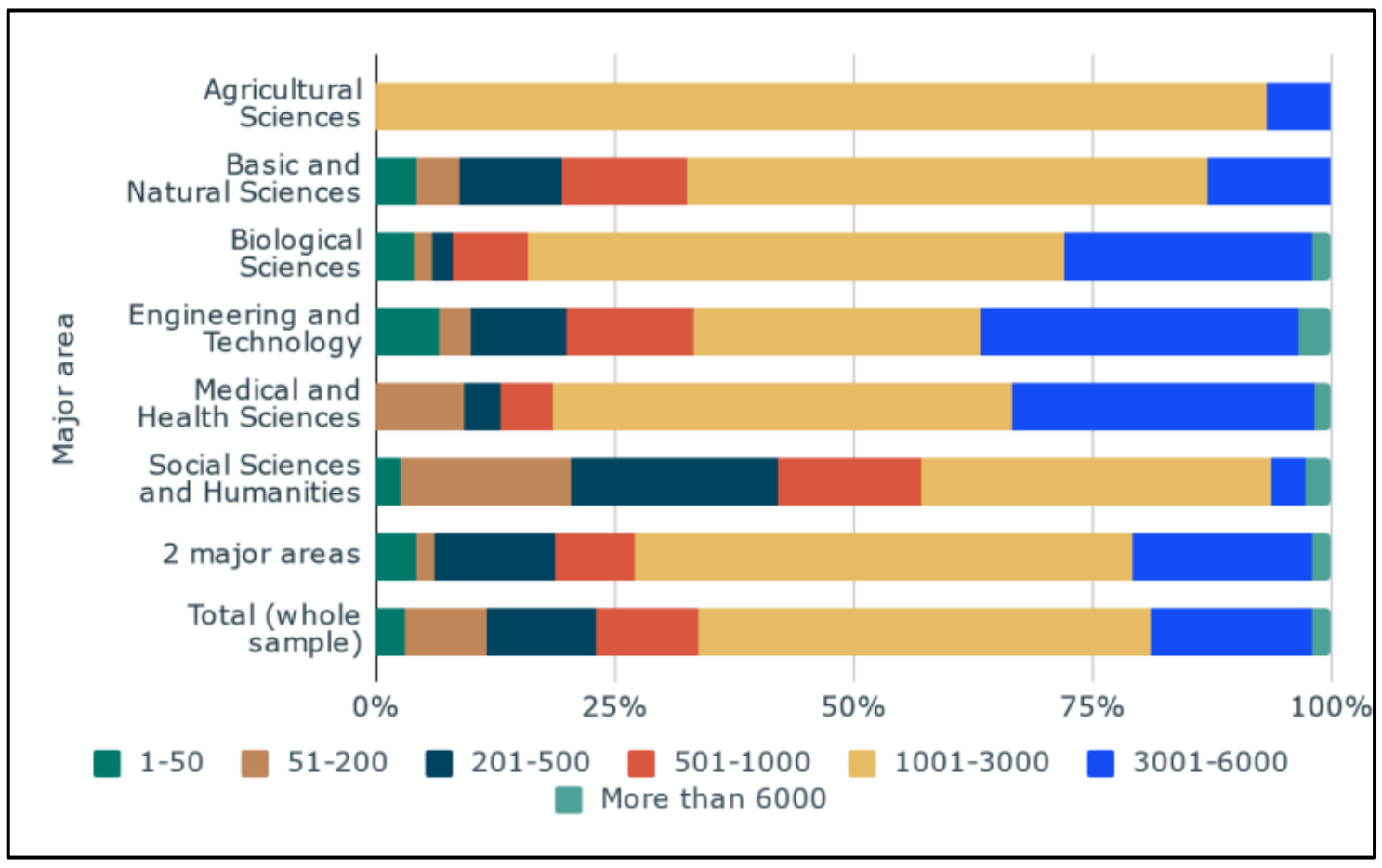

Source: Authors' elaboration based on data from APC surveys 2023 (GRIP project).

In Brazil, Mexico and Argentina the scientific field with the highest APCs is Medical and Health Sciences, followed by the Biological Sciences. In South Africa, researchers in Engineering and Technology paid the highest APCs, although the Medical and Health Sciences and Biological Sciences follow close behind. In all cases, Social Sciences and Humanities researchers reported paying the lowest APCs; but, for South Africa, the proportion of researchers in the Social Sciences and Humanities paying over USD 1,000 is the highest. Again, the availability of institutional funding and/or of read-and-publish deals seem to explain the extension of South African researchers in the Social Sciences and Humanities that publish in APC journals.

## APC and international collaboration

The final question in the APC section of the survey asked whether researchers collaborated with others because those researchers were able to pay APCs (see Table 12).

*Table 12. Researchers who collaborated with other authors that could afford APC payments, by country. N=8,289*

| APC payment | Argentina | Mexico | Brazil | South Africa |
|---|---|---|---|---|
| Have collaborated with authors who can afford APCs | 22,5% | 24,5% | 33,5% | 12,1% |
| Have collaborated with authors who can afford APCs | 74,7% | 73,0% | 63,2% | 87,0% |
| Prefer not to say | 2,7% | 2,5% | 3,3% | 1,0% |
| **Total** | **100%** | **100%** | **100%** | **100%** |

Source: Authors' elaborations based on data from APC surveys 2023 (GRIP project).

The selection of co-authors for the purposes of paying APCs is particularly important for Argentina (22,5%), Mexico (24,5%) and Brazil (33,5%) compared with South African researchers (12,1%). The distributions are similar when the data are analyzed by gender, age, research category, or type of research project.

## Conclusions

This paper set out to provide an empirical overview of the payment of APCs by a nationally rated group of researchers in four countries: Argentina, Brazil, Mexico and South Africa. It paid specific attention to the academic age of researchers, gender, scientific field, and collaboration. In terms of age, the findings show that the CONICET researchers in Argentina have a larger young cohort of researchers compared to the other three countries. However, it is mainly older researchers who publish in APC journals. Argentina is also a country in which researchers rely mostly on personal funds, APC waivers, and international authors to pay for APCs. This suggests the absence of support at the institutional and national levels for researchers to pay APCs. This may also explain the minor incidence of APC payments among young researchers. Conversely, Brazil, South Africa, and Mexico have a significant share of older researchers in the sample, but still, younger researchers publish more in APC journals.

This study has found that the most common range for maximum APC payments across the four surveyed countries is USD 1,001 to 3,000. In South Africa, researchers more frequently reported APC payments above USD 3,000. Generally, the higher the percentage of researchers who pay APCs from institutional or research funds, the higher the average APC paid in that country. Conversely, relying on APC waivers or personal funds leads to lower APC payments. This is especially notable among Argentinian researchers.

Regarding scientific disciplines, in Brazil, Mexico, and Argentina, researchers in Medical and Health Sciences pay the highest APCs, followed by Biological Sciences. However, in South Africa, engineering and technology researchers pay the highest APCs, with medical and health sciences and biological sciences closely following. Social Sciences and Humanities researchers consistently report paying the lowest APCs. As mentioned, however, South African researchers in these fields have a higher proportion of researchers paying over USD 1,000.

International collaboration with other authors for APC payment is more common in Argentina, Mexico and Brazil when compared to South Africa. The selection of co-authors that could afford the APC payments is particularly more important in those three countries. Gender, age, research category, and project type do not significantly alter this pattern.

A significant finding regarding the researchers' gender is the higher incidence of APC journals among men than among women, which is found in virtually all segments defined by major disciplinary fields, career categories, and both biological and academic age. In further investigations, special attention should be paid to possible correlation between gender, authorship position, and APC. More precisely, it is necessary to explore if men are more likely to be first, last or corresponding authors than women in these national cases in each disciplinary field, and to correlate author position and APC payments.

In terms of the scientific field, the South African sample is unique in the sense that it comprises a relatively larger proportion of researchers from the Social Sciences and Humanities. These scholars publish less frequently in APC journals, and yet, despite their relatively large representation in the national sample, South Africa publishes the highest proportion of APC articles of the four countries. This may be related to some factors regarding South African scenario, such as (1) researchers in that country have greater access to institutional open access publication funds, (2) its unique system of financial rewards for publications, and/or (3) its number of "read and publish" agreements.

The GRIP team is currently writing other papers to explore in greater depth specific issues related to the results of the survey in terms of payment of APCs and their effects on the scholarly publishing system, building upon the baseline data provided in this paper. These include a) a study on section 2 of the survey, regarding the perceptions of the researchers towards what a prestigious journal is, and its relationship with each national journal classification systems, b) a report on Section 4, investigating the levels of knowledge about the Open Access alternatives or business models and their effects in: publishing practices, along with the role played in each country by national regulations and available open infrastructures, and finally c) a study on Section 5, focused on perceptions, challenges, and ethical position takings regarding paying to publish.

We hope that this paper will not only serve as an important reference for the GRIP survey but also for all scholars exploring the emergence of APCs and its effects on the scholarly publishing system, as well as on the ensuant practices of researchers, particularly those who ply their trade in the relatively resource-scarce scientific periphery.

## Annexure: The survey questionnaire

The following questionnaire is the South African version of the online survey instrument. The surveys used in Mexico, Argentina, and Brazil have the same structure, with minor variations in Section 1 to accommodate language and national differences in terms of institutional affiliation types.

### SECTION 1: General information

#### 1.01. Are you currently a researcher recognized at the NRF?*

Please choose only one of the following:
- ☐ Yes
- ☐ No

#### 1.02. In which major scientific area would you include your main research activities?*

Please select from 1 to 2 answers:
- ☐ Agricultural Sciences
- ☐ Biological Sciences
- ☐ Engineering and Technology
- ☐ Basic and Natural Sciences
- ☐ Medical and Health Sciences
- ☐ Social Sciences and Humanities

#### 1.03. Agricultural Sciences. In which discipline would you include your main research activities?*

Please select from 1 to 2 answers.
- ☐ Agricultural Engineering
- ☐ Agronomy
- ☐ Animal Science
- ☐ Fisheries Resources and Fisheries Engineering
- ☐ Food Science and Technology
- ☐ Forest Resources and Forest Engineering
- ☐ Veterinary Medicine
- ☐ Other:

#### 1.03. Biological Sciences. In which discipline would you include your main research activities?*

Please select from 1 to 2 answers.
- ☐ Biochemistry
- ☐ Biophysics
- ☐ Botany
- ☐ Ecology
- ☐ General Biology
- ☐ Genetics
- ☐ Immunology
- ☐ Microbiology
- ☐ Morphology
- ☐ Parasitology
- ☐ Pharmacology
- ☐ Physiology
- ☐ Zoology
- ☐ Other:

#### 1.03. Engineering and Technology. In which discipline would you include your main research activities?*

Please select from 1 to 2 answers.
- ☐ Aerospace Engineering
- ☐ Biomedical Engineering
- ☐ Chemical Engineering
- ☐ Civil Engineering
- ☐ Electrical Engineering
- ☐ Materials and Metallurgical Engineering
- ☐ Mechanical Engineering

- ☐ Mining Engineering
- ☐ Naval and Oceanic Engineering
- ☐ Nuclear Engineering
- ☐ Production Engineering
- ☐ Sanitary Engineering
- ☐ Transport Engineering
- ☐ Other:

### 1.03. Basic and Natural Sciences. In which discipline would you include your main research activities?*

Please select from 1 to 2 answers.

- ☐ Astronomy
- ☐ Chemistry
- ☐ Computer Science
- ☐ Geosciences
- ☐ Mathematics
- ☐ Oceanography
- ☐ Probability and Statistics
- ☐ Physics
- ☐ Other:

### 1.03. Medical and Health Sciences. In which discipline would you include your main research activities?*

Please select from 1 to 2 answers.

- ☐ Clinical Medicine
- ☐ Dentistry
- ☐ Deontology
- ☐ Forensic Medicine
- ☐ Maternal and Child Health
- ☐ Medical Radiology
- ☐ Nursing
- ☐ Nutrition
- ☐ Pathological Anatomy and Clinical Pathology
- ☐ Pharmacy
- ☐ Phonoaudiology
- ☐ Physical Education
- ☐ Physiotherapy and Occupational Therapy
- ☐ Psychiatry
- ☐ Public Health
- ☐ Surgery
- ☐ Other:

### 1.03. Social Sciences & Humanities. In which discipline would you include your main research activities?*

Please select from 1 to 2 answers.

- ☐ Administration
- ☐ Anthropology
- ☐ Archaeology
- ☐ Architecture and Urbanism
- ☐ Arts
- ☐ Communication
- ☐ Demography
- ☐ Economics
- ☐ Education
- ☐ Geography
- ☐ History
- ☐ Home Economics
- ☐ Industrial Design
- ☐ Information Science
- ☐ Languages
- ☐ Linguistics
- ☐ Law
- ☐ Museology
- ☐ Philosophy
- ☐ Political Science
- ☐ Psychology
- ☐ Social Work
- ☐ Sociology
- ☐ Theology
- ☐ Tourism
- ☐ Urban and Regional Planning
- ☐ Other

### 1.04. Please indicate your current NRF rating.*

Please choose only one of the following:

- ○ A1
- ○ A2
- ○ B1
- ○ B2
- ○ B3
- ○ C1
- ○ C2
- ○ C3
- ○ P
- ○ Y1
- ○ Y2
- ○ Other

### 1.05.1. What is your main institutional affiliation as a researcher? (Please write out the full name of the institution) *

Please write your answer here:

### 1.05.2. In which city is your institution located

Please write your answer here:

### 1.06. What is your highest academic degree?*

Please choose only one of the following:

- ○ PhD or above
- ○ Master's degree
- ○ Other

### 1.07. In what year did you complete your highest academic degree?*

Please write your answer here:

### 1.08. What is your year of birth?*

Please write your answer here:

### 1.09. What is your gender?*

Please choose only one of the following:

- ○ Male
- ○ Female
- ○ Non-binary
- ○ I prefer not to say
- ○ Other

### 1.10. What type of research projects have you participated in during the past 5 years?*

Please choose all that apply:

- ☐ National public funded project(s)
- ☐ International public funded project(s)
- ☐ National private funded project(s)
- ☐ International private funded project(s)
- ☐ National ONG (Non-Governmental National Organizations) funded project(s)
- ☐ International ONG (Non-Governmental International Organizations) funded project(s)
- ☐ Other:

## SECTION 2: Publication of results and evaluation of journals

### 2.01. How important are the following means for disseminating your work?*

Please choose the appropriate response for each item:

| *Options* | *Very important* | *Somewhat important* | *Not important* | *Not applicable to my area of study* |
|---|---|---|---|---|
| Community engagement activities | ○ | ○ | ○ | ○ |
| Creative works | ○ | ○ | ○ | ○ |
| Presenting at academic conferences | ○ | ○ | ○ | ○ |
| Technology transfer | ○ | ○ | ○ | ○ |
| Reports, briefs, occasional papers | ○ | ○ | ○ | ○ |
| Publication of scholarly books | ○ | ○ | ○ | ○ |
| Publication in conference proceedings | ○ | ○ | ○ | ○ |
| Publication in scientific journals | ○ | ○ | ○ | ○ |
| Patent registration | ○ | ○ | ○ | ○ |
| Publication of book chapters | ○ | ○ | ○ | ○ |
| Publication in professional periodicals | ○ | ○ | ○ | ○ |
| Publication in popular media | ○ | ○ | ○ | ○ |

### 2.02. What factors do you consider when selecting a scientific journal to publish in? *

Please select from 1 to 4 answers.

- ☐ The ease with which I will be able to publish in the journal
- ☐ The scientific domain of the journal and its relevance to my field of study
- ☐ The database in which it is indexed (e.g. Scopus, Web of Science, Scielo)
- ☐ The number of citations or impact factor of the journal
- ☐ The immediate open access publication of the article in the journal
- ☐ The cost of publication (e.g., APCs, colour charges, etc.)
- ☐ Value of the journal in the evaluation system
- ☐ Relevance to the local agenda (pressing problems in my region or country)
- ☐ Peer review (type, e.g. double blind vs single blind vs open)
- ☐ Time taken to publish (including peer review time)
- ☐ Recommendations by peers and/or mentors
- ☐ The publisher of the journal
- ☐ The readership of the journal
- ☐ Rejection rate of the journal
- ☐ The language of the journal
- ☐ The members of journal's the editorial board
- ☐ Other:

### 2.02.1. You indicated that the value of a journal in the evaluation system is an important consideration when selecting a journal to publish in. Could you be more specific about which evaluation system?*

Please choose only one of the following:

- ○ Institutional evaluation (e.g. the evaluation system of your university for promotion or tenure)
- ○ National evaluation (e.g. national funding ratings or categorizations by science councils)
- ○ International evaluation (e.g. the evaluation for international funds or international rankings)
- ○ Other:

### 2.03. In your opinion, what are the main indicators of a prestigious journal in your field?*

Please choose all that apply:

- ☐ Indexing in international databases
- ☐ High impact factor
- ☐ Publication of articles with novel results
- ☐ Frequent citation among colleagues
- ☐ International status
- ☐ Publication of articles by recognised or preeminent scholars in the field
- ☐ Reputation of the publisher, scientific society or university
- ☐ Rigorous peer review
- ☐ Recognised scholars as editors of the journal or members of the editorial board

- ☐ Open access (no fees)
- ☐ Open access (with APCs)
- ☐ Other:

## SECTION 3: Types of open access

### 3.01. Do you know what Open Access is?*

Please choose only one of the following:

- ○ Yes, I do
- ○ No, I don't
- ○ Other

### 3.01.1. Do you know the following types of Open Access?*

Please choose all that apply:

- ☐ Green
- ☐ Gold
- ☐ Diamond
- ☐ Hybrid
- ☐ Other

Definitions:

- Green OA: making a version of the manuscript freely available in a repository.
- Gold OA: making the final version of the manuscript freely available immediately upon publication by the publisher subject to the payment of APCs.
- Diamond OA: a journal that publishes the articles immediately in open access and that does not charge any fee, both for reading or publishing.
- Hybrid: a journal that publishes both subscription and open access articles, where authors are given the option to publish open access if they pay APCs.

### 3.02. Have you ever published your research in an open access journal?*

Please choose only one of the following:

- ○ Yes, I do
- ○ No, I don't
- ○ I do not know

Definition: Open Access (OA) - "By 'open access' we mean the free availability of the scientific literature on the public internet, permitting any users to read, download, copy, distribute, print, search, or link to the full texts of these articles, crawl them for indexing, pass them as data to software, or use them for any other lawful purpose, without financial, legal, or technical barriers other than those inseparable from gaining access to the internet itself. The only constraint on reproduction and distribution, and the only role for copyright in this domain, should be to give authors control over the integrity of their work and the right to be properly acknowledged and cited." (Budapest Open Access Initiative, 2002).

### 3.02.1. In what type of open access journal have you published?*

Please choose all that apply:

- ☐ I have published in a gold open access journal (a journal that charges APCs to publish articles in open access)
- ☐ I have published in a diamond open access journal (a journal that doesn't charge APCs to publish articles in open access and there is no author fee)
- ☐ I have published in a hybrid journal (a journal that publishes some articles in open access by charging authors APC fees, while other articles remain behind a paywall)

- ☐ I have published a preprint (a manuscript that has not yet been submitted/subjected to a formal peer review and is distributed in advance to receive feedback about research from colleagues in the field or other people interested in the topic)
- ☐ I do not know the type of Open Access
- ☐ Other:

### 3.02.2. How many papers have you published in the last 5 years?*

Please choose only one of the following:

- o 1-5
- o 6-10
- o 11-20
- o 21 or more

### 3.02.3. How many of these papers were published in Open Access journals?*

Please choose only one of the following:

- o 1-5
- o 6-10
- o 11-20
- o 21 or more
- o No, I have not published any articles in Open Access

### 3.02.4. Please specify the reasons why you have not published in open access journals?*

Please choose all that apply:

- ☐ I do not know of any suitable open access journals
- ☐ I have not been able to afford the publication fees
- ☐ I refuse to pay to publish my research results in open access journals
- ☐ In my field open access journals are not listed in rankings and/or are not well rated/ valued in evaluation systems
- ☐ In my field, open access journals have a poor reputation
- ☐ I am afraid that these open access journals are predatory
- ☐ I chose to publish in another type of publication (e.g. a blog, new media, etc.)
- ☐ Other:

### 3.03. Have you ever published in a Diamond open access journal?*

Please choose only one of the following:

- o Yes
- o No
- o Other

### 3.03.01. Why have you not published in a diamond open access journal?*

Please select from 1 to 2 answers.

- ☐ These journals do not exist in my field
- ☐ I do not know any diamond journals
- ☐ I have not had an opportunity to publish in diamond journals
- ☐ Diamond journals are not recognized by evalutation systems
- ☐ In my field, diamond journals have a bad reputation
- ☐ In my country, diamond journals have a bad reputation

- ☐ Because they are not covered by international databases
- ☐ I am afraid of diamond journals being predatory
- ☐ Other:

### 3.03.02. What is the main reason to choose to publish in a diamond open access journal?*

Please choose only one of the following:

- ○ I refuse to pay APCs
- ○ I did not have enough funding to pay for publication fees
- ○ There are prestigious diamond journals in my field
- ○ Diamond journals are means to reach a wider audience
- ○ Other

### 3.03.03. Could you explain why you refuse to pay APCs?*

Please write your answer here:

## SECTION 4: Publishing in APC journals and funding

### 4. In the last five years, have you published in a journal that charges APCs?*

Please choose only one of the following:

- ○ Yes
- ○ No
- ○ I don't know

### 4.01.1 How were these APCs more frequently paid?*

Please select from 1 to 3 answers. Drag your selection from the box on the left-hand side to the empty box on the right-hand side

- ☐ Total or partial payment with my research project funds
- ☐ I had the option not to pay an APC and the article had restricted access
- ☐ Total or partial payment with personal funds
- ☐ The payment was made through a specific APC fund of my university/
- ☐ Application for an APC waiver (either a discount or a full exemption)
- ☐ The payment was made with funds from co-authors of other institutions
- ☐ The payment was made with funds from co-authors of foreign institutions
- ☐ Other
- ☐ I do not know

### 4.01.2. What sources and how frequently were they used to pay for full or partial APCs?*

Please choose the appropriate response for each item:

| *Options* | *Never* | *Once* | *More than once* |
|---|---|---|---|
| Payment with institutional/university research project funds | ○ | ○ | ○ |
| Payment with my own (personal) funds or salary | ○ | ○ | ○ |
| Payment with national/ federal research funds | ○ | ○ | ○ |
| Payment with international research funds | ○ | ○ | ○ |

### 4.01.3. What is the maximum APC you have paid to publish an article?*

Please choose only one of the following:

- ○ US$ 1-50
- ○ US$ 51-200
- ○ US$ 201-500
- ○ US$ 501-1,000

- US$ 1,001-3,000
- US$ 3,001-6,000
- More than US$ 6,000
- Don't know
- Prefer not to say

### 4.02. Have you ever collaborated with certain authors because they were able to pay APCs?*

Please choose **only one** of the following:

- Yes
- No
- Prefer not to say

## SECTION 5: Final comments

### 5.01. With which of the following statements do you agree most?*

Please choose only one of the following:

- I do not consider it relevant whether a journal is open access when I submit my articles for publishing
- I only consider it relevant to publish in high-impact journals regardless of its open access policies
- I consider it relevant to publish my articles in diamond open-access journals
- I consider it relevant to publish my articles in prestigious journals regardless of the Impact Factor
- None of the above

### 5.02. What are the main challenges regarding APCs?*

Please choose all that apply:

- ☐ APCs are not funded through Institutional/governmental agreements with the major publishers
- ☐ Most journals in my field charge APCs
- ☐ Most of the high-quality journals in my field charge APCs
- ☐ The amount charged for APCs has increased considerably
- ☐ It is difficult to make payments in foreign currency from my country
- ☐ It is difficult to make APC payments due to university/organizational procurement rules
- ☐ The funding in my country available to cover APC payments is insufficient
- ☐ The funding in my field available to cover APC payments is insufficient
- ☐ APC payments can be covered by colleagues from abroad, but this implies losing prominence in publications
- ☐ I do not agree with APC (commercial publishing)
- ☐ I don't see any challenges regarding APCs
- ☐ Other:

### 5.03. Do you plan to use research project funds to pay for APCs in the future?*

Please choose only one of the following:

- Yes
- No
- Other

### 5.04. With which of the following statements do you agree most?

Please choose only one of the following:
Article Processing Charges…

- Are likely to foster a better, more selective, publishing process
- Are the main problem for getting my work published open access
- Are not currently a problem to get my work published open access
- Are likely to represent a problem in the next few years
- Are likely to improve the quality of journal articles
- Other

**5.05. Do you think your publishing practices have changed due the journal's shift to open access?**

Please write your answer here:

**5.06. Any final comments or recommendations regarding the payment of APCs to publish?**

Please write your answer here:

**Would you like to receive an electronic copy of the final report of this research project?**

Please choose only one of the following:

- Yes
- No

Thank you! We appreciate your time. If you are interested in this survey and want to receive the final report and its results, please contact us at the email address below. Also, if you are interested in an interview about this topic, please let us know. apc.grip@listes.u-paris.fr (mailto:apc.grip@listes.u-paris.fr)

Submit your survey.

Thank you for completing this survey.